\documentclass[fleqn,usenatbib]{rasti}

\usepackage[T1]{fontenc}

\DeclareRobustCommand{\VAN}[3]{#2}
\let\VANthebibliography\thebibliography
\def\thebibliography{\DeclareRobustCommand{\VAN}[3]{##3}\VANthebibliography}

\usepackage{graphicx}	
\usepackage{amsmath}	
\usepackage{bm}
\usepackage{soul}
\usepackage{xcolor}
\usepackage{xparse}
\usepackage{ dsfont }
\usepackage{IEEEtrantools}

\newcommand{\coh}{\gamma^2}

\newcommand{\cohhat}{\hat{\gamma}^2}
\NewDocumentCommand{\edit}{+m}{\textcolor{black}{#1}}

\newenvironment{edit-sec}{%
  \color{black}%
}{%
  \normalcolor%
}

\title[Correlated Time Series for X-ray Astronomy]{Generation of Correlated Time Series for X-ray Astronomy Applications}

\author[S. R. Larner et al.]{
Seth Rowan Larner,$^{1}$\thanks{E-mail: larner@wustl.edu (SRL)}
Michael A. Nowak,$^{1,2}$
and J\"orn Wilms$^{3}$
\\
$^{1}$Department of Physics, Washington University in St.~Louis, St.~Louis, MO 63130, USA\\
$^{2}$McDonnell Center for the Space Sciences, Washington University in St.~Louis, St.~Louis, MO 63130, USA\\
$^{3}$Dr.\ Karl Remeis-Observatory and Erlangen Centre for Astroparticle Physics, Universit\"at Erlangen-N\"urnberg, Sternwartstr.~7, 96049\\ Bamberg, Germany
}

\date{Accepted XXX. Received YYY; in original form ZZZ}

\pubyear{\the\year{}}

\begin{document}
\label{firstpage}
\pagerange{\pageref{firstpage}--\pageref{lastpage}}
\maketitle

\begin{abstract}
Cross-spectral methods have become essential for studying accretion physics in X-ray binaries and active galactic nuclei, where coherence and phase lag measurements constrain physical models and reveal variability components invisible in power spectra alone. Recent multi-Lorentzian fitting techniques have uncovered new quasi-periodic features through joint analysis of power spectra and cross-spectra, but testing these methods requires synthetic data with realistic statistical properties. We present an algorithm for generating pairs of time series with arbitrary power spectra, coherence functions, and phase lag profiles. The method extends existing methods
by decomposing the dependent time series into coherent and incoherent components, where the coherent part is constructed through a complex transfer function applied to a reference series. We derive the transfer function and normalization required to preserve target spectral shapes while achieving specified cross-spectral properties. We obtain approximate analytic expressions for the variance of coherence and phase lag estimators and construct the approximate covariance matrix relating these quantities to the underlying power and cross-spectra. When fitting models jointly to power and cross-spectra, these correlations must be incorporated into the likelihood. We apply the method to a two-Lorentzian model with component-specific phase lags and demonstrate close agreement between input models and generated power spectra, coherence function, and phase lag profile across four decades in frequency.
\end{abstract}




\section{Introduction}
\label{sec:intro}

Fourier-domain techniques are fundamental to the study of variability in accreting compact objects. X-ray binaries (XRBs) and active galactic nuclei (AGN) are observed to be highly variable on timescales of ms to years. The structure of this variability encodes information about the geometry and physics of accretion flows which is otherwise inaccessible through spectral analysis alone \citep{vanderklis1989,uttley2014}.
Commonly used Fourier techniques include power spectral densities (PSDs), cross power spectral densities (CSDs), and the frequency-dependent phase lags, $\phi(\nu)$, and coherence, $\coh(\nu)$.
The cross-spectrum $C(\nu) = F_1^*(\nu) F_2(\nu)$ of two simultaneously measured light curves is a complex quantity encoding both the degree of correlation and the phase relationship between variability in the two bands.
The coherence function $\coh$, whose use in X-ray astronomy was highlighted by \citet{vaughan1997}, is the squared magnitude of the normalized cross-spectrum. It quantifies the degree of linear correlation as a function of Fourier frequency, ranging between zero (completely uncorrelated) and unity (perfectly correlated). The argument of the cross-spectrum gives the phase lag $\phi(\nu)$ between the two signals.
When combined with coherence information, phase lags reveal the transfer function relating variability in one band to another.
These techniques allow for comparison of variability processes across energy bands and have the promise of constraining models of accretion disk-corona systems \citep[e.g.,][]{kara2019,demarco2021} and might establish the existence of independent variability components in the X-ray band \citep{nowak1999b}.

These diagnostic tools have proven powerful for testing physical models. \citet{vaughan1997} demonstrated that the near-unity coherence observed between soft and hard X-rays in Cyg~X-1 and GX~339$-$4 rules out a broad class of models invoking spatially extended fluctuating emission regions, thermal flares, or overlapping shot noise. Subsequent work established that drops in coherence at specific frequencies could indicate the superposition of physically distinct variability components \citep{nowak1999b}.

The phase lag spectrum typically shows harder emission lagging behind soft (``hard lags'') at low frequencies and soft emission lagging behind hard (``soft lags'') at higher frequencies \citep[e.g.,][]{miyamoto1988,nowak1999a,pottschmidt2003,grinberg2014}.
Hard lags have sometimes been attributed to the inward propagation of mass accretion rate fluctuations \citep{arevalo2006}, while soft lags at higher frequencies have been interpreted as tracing X-ray reverberation from the inner accretion disc \citep{uttley2014,kara2016}.

Recent work has demonstrated the power of jointly modeling the full cross-spectral information. \citet{mendez2024}
introduced a technique for simultaneously fitting the PSD and the real and imaginary parts of the cross-spectrum of XRBs with multi-Lorentzian models under the assumption that individual Lorentzian components are each internally coherent across energy bands but mutually incoherent with one another.
This framework has revealed ``hidden'' variability features. These features appear to be undetectable in the PSD alone but are prominent in the cross-spectrum. They have been interpreted as quasi-periodic oscillations (QPOs), similar to the low frequency QPOs often found in XRBs \citep{konig2024,mendez2024,fogantini2025}.
The success of these cross-spectrum-first approaches reveals the richness of information encoded in the coherence and phase lag, beyond what is accessible from power spectra alone.

Simulating realistic time series with prescribed cross-spectral properties is essential for validating these analysis techniques, testing model predictions, and understanding the statistical behavior of Fourier-domain estimators. The standard method for generating non-deterministic time series with a prescribed power spectrum was introduced by \citet{timmer1995}. Briefly, the method constructs a Fourier transform by drawing the real and imaginary components at each frequency from Gaussian distributions with variance proportional to the target PSD, then inverse transforms to obtain the time-domain realization. This approach correctly reproduces the stochastic scatter in PSD estimates and has become ubiquitous in X-ray timing work, applied extensively to, e.g., characterize AGN variability \citep{uttley2002}, test lag recovery techniques \citep{zoghbi2013}, and model propagating fluctuations \citep{ingram2013}.

However, even when generated from the same target PSD, light curves generated with the \citet{timmer1995} method are statistically independent. By construction, when averaging over many segments, two realizations will exhibit zero coherence and net phase lag. Consequently, these realizations are not suitable for analysis of correlated light curves with non-trivial phase relations. Previous work has extended the \citet{timmer1995} method to produce non-Gaussian distributed time series which more closely match the observed properties of time series from X-ray binaries and AGN \citep{emmanoulopoulos2013}.

We present a method for simulating pairs of time series that can have different power spectral shapes while exhibiting arbitrary frequency-dependent coherence and phase lag profiles. The approach generalizes the \citet{timmer1995} algorithm by constructing the Fourier transform of a ``dependent'' time series as a weighted superposition of a coherent component (derived from a ``reference'' time series modulated by a complex linear transfer function) and an incoherent component. We derive the relationship between the coherence and phase lag and the transfer function and demonstrate that proper normalization preserves the target PSDs in both time series.

In Section~\ref{sec:definitions}, we establish notation and review the definitions of the cross-spectrum, coherence, and related quantities. In Section~\ref{sec:derivations}, we derive the key equations relating coherence and phase lag to the transfer function and establish the normalization required to preserve the power spectrum and provide the step-by-step algorithm. In Section~\ref{sec:variance}, we derive an analytic expression for the variance in the estimated coherence and phase lag as a function of the number of averaged segments, providing guidance on how much data is required to achieve a given precision.
In Section~\ref{sec:results}, we demonstrate the method on representative coherence and phase lag profiles and verify the analytic predictions against Monte Carlo simulations. We summarize our conclusions in Section~\ref{sec:conclusions}.

\section{Mathematical Background and Formalism}
\label{sec:definitions}

Consider two noiseless time series consisting of $K$ evenly-spaced and aligned samples with no gaps, \(\textbf{x} = \{x_i\}_{i=0}^{K-1} \) and \(\textbf{y} = \{y_i\}_{i=0}^{K-1} \). We form the discrete Fourier transforms of these via
\begin{equation}
    X_j\equiv\sum_{k = \edit{0}}^{\edit{K-1}} x_k \ e^{\frac{-2 \pi i j k}{K}} \quad\mbox{and}\quad
     Y_j \equiv \sum_{k = \edit{0}}^{\edit{K-1}} y_k \ e^{\frac{-2 \pi i j k}{K}}
\end{equation}
where the discrete Fourier frequency for time series of duration $T$ is given by \(\nu_j = \frac{j}{T}, j = 0, 1, \ldots, K-1\)\footnote[1]{\edit{For real-valued time series \textbf{x} and \textbf{y}, one is free to express only K/2 + 1 independent Fourier coefficients. See Appendix~\ref{app:spectral-leakage} for details.}}. Using these, we define the PSDs,
\begin{equation}
    P_X(\nu_j) \equiv E[X_j X_j^*] \quad \mbox{and}\quad
     P_Y(\nu_j) \equiv E[Y_j Y_j^*]
\end{equation}
and the cross spectrum,
\begin{equation}
    C(\nu_j) \equiv E[X_j^* Y_j]
\end{equation}
where ${}^*$ denotes a complex conjugate. We adopt the following notational convention to distinguish between mathematical expectation and finite sample averaging: we use $E[\cdot]$ to denote the expectation value of a random quantity (the theoretical or limiting value), and we use $\langle \cdot \rangle$ to denote averaging over a finite number of data segments. Quantities estimated from finite data are denoted with a hat ($\hat{\bullet}$) to distinguish them from their true underlying values. For example, by averaging $X_j X^*_j$ over $n$ segments, we calculate $\hat{P}_X = \langle X_j X^*_j \rangle$, which estimates the true value $P_X = E[X_j X^*_j]$. 

From the cross spectrum, we can derive the coherence, sometimes called the intrinsic coherence to distinguish it from the coherence of signals that have been modified by detector effects and time-binning (see Appendix~\ref{app:detector_coherence}) and Poisson noise (see equation \ref{eq:poisson_dilution}),
\begin{equation}
\label{eq:coherence}
    \coh(\nu_j) = \frac{|C(\nu_j)|^2}{P_X P_Y}
\end{equation}
and the phase lag,
\begin{equation} \label{eq:phase}
    \phi(\nu_j) = \arg[C(\nu_j)],
\end{equation}
choosing $\phi \in (-\pi, \pi]$ with positive phase indicating $Y$ leads $X$.

We will, in general, consider the time series \textbf{y} (the ``dependent'' time series) to be composed of two components, a component that is uncorrelated with \textbf{x} and a component that is completely linearly correlated with \textbf{x} (the ``reference'' time series)
\footnote[2]{This construction neglects any possible higher-order correlations between the two time series. In the most general case, \textbf{y} could be expressed as a Volterra series in \textbf{x}
\citep[see][for a mathematical discussion]{schetzen1980}.
Nonzero higher-order terms would each admit their own nonzero correlation functions analogous to the linear coherence function we consider here \citep{kim1979}. We are unaware of any consideration of these higher order terms in the context of XRB or AGN studies, though we note that bicoherence -- which probes nonlinear phase coupling within a single time series -- has been used to study the multiplicative nature of accretion variability \citep{arur2022,maccarone2002}.}.
One could alternatively imagine \textbf{x} and \textbf{y} both being related to some shared signal, with each containing their own uncorrelated component. For the purposes of calculating the power and cross spectra, however, it is sufficient to absorb all uncorrelated signal into \textbf{y}
\footnote[3]{We direct interested readers to the discussion in \citet{nathan2026}. We have additionally provided verification that writing an uncorrelated term in both time series reduces to this simpler construction when calculating the expected values of the power and cross spectra along with their variances and covariances (the results of Sections \ref{sec:derivations} and \ref{sec:variance} and Appendix~\ref{app:covariance_structure}) in the form of a Wolfram Mathematica notebook \citep{Mathematica} publicly available on Github at \texttt{https://github.com/SethHtes/RASTI-scripts}.}.

With this decomposition, we write
\begin{equation}
    \begin{aligned}
    y_i &= y_{u,i} + y_{c,i} \\
    &= y_{u,i} + [r(t_i, E_1, E_2) * x]_{i},
    \end{aligned}
\end{equation}
where we choose to express the correlated component of \textbf{y} as the convolution of \textbf{x} and a transfer function $r$ which is a function of time $t$ and reference and dependent energy bands $E_1$ and $E_2$. The Fourier transform of the dependent time series is then given by
\begin{equation}
\label{eq:dependent_fourier_intro}
    Y_j = Y_{u,j} + R(\nu_j, E_1, E_2) X_{j}.
\end{equation}
Hereafter, we will neglect the subscript notation indicating discrete time series and Fourier transforms, for convenience. We will also consider only two time series at a time, allowing us to neglect the energy-dependence of $R$.


\subsection{Bias of the Coherence Estimator}
\label{sec:bias}

The estimator $\hat{\gamma}^2$ of the coherence function defined by equation~(\ref{eq:coherence}) is a biased estimator of the true coherence $\gamma^2$. \citet{NuttallCarter1976} derived the exact expected value of this estimator when averaging over $n$ independent segments
\begin{equation}
    E[\hat{\gamma}^2] = \frac{1}{n} + \frac{n-1}{n+1} \gamma^2 \, {}_2F_1(1, 1; n+2; \gamma^2)
\end{equation}
where ${}_2F_1$ is the Gaussian hypergeometric function. The bias $E[\hat{\gamma}^2] - \gamma^2$ is strictly positive for $\gamma^2 < 1$, meaning the estimator systematically overestimates coherence. This effect is most pronounced at low coherence values and small $n$.

For practical purposes, \citet{NuttallCarter1976} provide the approximation
\begin{equation}
    \label{eq:bias}
    E[\hat{\gamma}^2] - \gamma^2 \approx \frac{(1 - \gamma^2)^2}{n}
\end{equation}
which captures the leading-order behavior. The bias vanishes at $\gamma^2 = 1$ (perfect coherence) and reaches its maximum of $1/n$ at $\gamma^2 = 0$ (complete incoherence), where the expected value of the estimator is simply $1/n$. This bias has implications for interpreting measured coherence values, particularly in observations where limited segment counts preclude large $n$, as discussed by \cite{vaughan1997,nowak1999a,ingram2019}.

\section{Generating Correlated Time Series}
\label{sec:derivations}

Following \citet{timmer1995}, we generate the Fourier transform of the reference time series,
\begin{equation}
\label{eq:X}
    X = \left (\frac{P_X}{2} \right )^{1/2} (A_r + i B_r),
\end{equation}
where $P_X = E[X X^*]$ is the target power spectrum of the reference time series and the subscript $r$ denotes a realization of an independent normal random variable $A_r, B_r \sim \mathcal{N}(0,1)$.
Using equation~(\ref{eq:dependent_fourier_intro}), we write the Fourier transform of the dependent time series,
\begin{equation}
\label{eq:Y}
\begin{aligned}
    Y &= Y_{u} + R X \\
     &= \mathcal{K} (H_r + iJ_r) + R \left ({\frac{P_X}{2}} \right )^{1/2} (A_r + iB_r).
\end{aligned}
\end{equation}
Similar to the reference Fourier transform, we write the uncorrelated part of the dependent Fourier transform as the sum of random variables. Here, we leave the normalization constant $\mathcal{K}$ unspecified to preserve the target power spectral shape of the dependent time series $P_Y = E[Y Y^*]$.

\subsection{Derivation of the Normalization Constant}
\label{sec:normalization}
We wish to enforce a desired spectral shape ($P_Y$) on the dependent time series. We derive $\mathcal{K}$ by computing $E[YY^*]$ and setting it equal to $P_Y$. In the finite case, this ensures that $\hat{P}_Y = \langle Y Y^* \rangle $ approaches $P_Y$ with increasing averaging.

\begin{equation}
\begin{aligned}
    P_Y &= E[Y Y^*] \\
    &=  P_X |R|^2 + 2 |\mathcal{K}|^2
\end{aligned}
\end{equation}
where we have used $E[A_r] = E[B_r] = E[H_r] = E[J_r] = 0$ and $E[A_r^2] = E[B_r^2] = 1$. Thus, making the choice of $\mathcal{K}$ to be positive and real, we find
\begin{equation}
\label{eq:C}
    \mathcal{K} = \sqrt{\frac{P_Y - P_X |R|^2}{2}}.
\end{equation}

\subsection{Derivation of the Transfer Function}\label{sec:transfer-function}
We now turn to the transfer function $R(\nu)$ which modulates the reference Fourier transform when it is added to the dependent Fourier transform. This function is constrained by the target coherence function $\coh(\nu)$ and phase lag profile $\phi(\nu)$. Beginning with the coherence function,
\begin{equation}
\begin{aligned}
    \coh &= \frac{|E[X Y^*]|^2}{P_X P_Y }
    &= |R|^2 \frac{P_X}{P_Y} \\
    |R| &= \sqrt{\frac{P_Y \coh}{P_X}}.
\end{aligned}
\end{equation}
while for the phase lag profile,
\begin{equation}
\begin{aligned}
    \phi &= \arg[E[X^* Y]]
    &= \arg [R P_X]\\
    \arg [R] &= \phi.
\end{aligned}
\end{equation}
Thus, we can write
\begin{equation}
\label{eq:R}
    R = \sqrt{\frac{P_Y \coh(\nu)}{P_X}} e^{i \phi(\nu)}.
\end{equation}

\subsection{Summary of Method}
Having derived the necessary constants, we can fully define the proposed algorithm in the following steps:
\begin{enumerate}
    \item Choose a target power spectral shape $P_X$ for the reference time series  and $P_Y$ for the dependent time series.
    \item Choose a target coherence function $\coh$ and phase lag profile $\phi$.
    \item For each Fourier frequency, draw a random vector
    \begin{equation*}
        (A_r, B_r, H_r, J_r) \sim \mathcal{N}(\bm{0},\mathcal{I}_4)
    \end{equation*}
    where $\mathcal{I}_4$ is the $4\times4$ identity matrix.
    \item Calculate the constant $\mathcal{K}$ and the transfer function $R$ via equations (\ref{eq:C}) and (\ref{eq:R}), respectively.
    \item Generate the Fourier transforms of the reference and dependent time series via equations (\ref{eq:X}) and (\ref{eq:Y}).
    \item Obtain the time series by inverse Fourier transform of their respective Fourier transforms.
\end{enumerate}

The algorithm above parametrizes the simulation by $(P_X, P_Y, \gamma^2(\nu), \phi(\nu))$.
When a physical model instead specifies a response function $\mathcal{R}(\nu)$ directly, a more direct generation is possible. Draw reference Fourier
coefficients $X$ following \citet{timmer1995}, then form
\begin{equation*}
    Y = \mathcal{R}(\nu)\, X + \sqrt{\frac{P_N}{2}}\,(H_r + iJ_r),
\end{equation*}
where $P_N(\nu)$ is the incoherent noise power. The simulation is then parameterized by
$(P_X, \mathcal{R}, P_N)$ rather than $(P_X, P_Y, \gamma^2, \phi)$, and the cross-spectral
observables are outputs: $P_Y = |\mathcal{R}|^2 P_X + P_N$, $\phi = \arg\mathcal{R}$,
and $\gamma^2 = |\mathcal{R}|^2 P_X / P_Y$. The variance results of
Section~\ref{sec:variance} apply to both constructions. 

\section{Derivation of Variances and Covariances}
\label{sec:variance}

Having established a method for generating synthetic time series with prescribed cross-spectral properties, we now derive approximate analytic expressions for the variance of the estimated coherence and phase lag. These expressions quantify the statistical uncertainty in the measured coherence and phase lag as a function of the number of averaged segments and the true underlying coherence.
We do so using the delta method \citep{wolter2007}. 
For a scalar function $h(\bm{\hat{B}})$ of an estimator with true value $\bm{B}$ and covariance matrix $\Sigma$, the delta method gives \citep{wolter2007}
\begin{equation}
    \text{Var}(h(\bm{\hat{B}})) \approx \nabla h (\bm{B})^T \cdot \frac{\Sigma}{n} \cdot \nabla h (\bm{B}).
\end{equation}
Here, $\bm{\hat{B}} = (\hat{C}_r, \hat{C}_i, \hat{P}_X, \hat{P}_Y)$, where $C_r$ and $C_i$ are the real and imaginary parts of the cross-spectrum, and $h$ is either $\hat{\gamma}^2$ or $\hat{\phi}$.
The delta method approximates the true variance of the estimator under the assumption it is unbiased and normally-distributed. This holds in the limit of large $n$.

\subsection{Covariance Matrix}
\label{sec:covariance_matrix}
We can construct the covariance matrix $\Sigma$ by expressing each of $\hat{C}_r, \hat{C}_i, \hat{P}_X,$ and $\hat{P}_Y$ as quadratic forms of the random variable $\bm{\varepsilon} = (A_r, B_r, H_r, J_r)$. A quadratic form of an $n$-dimensional random variable $\bm{\varepsilon}$ is a scalar quantity that can be expressed as $\bm{\varepsilon}^T \Lambda \bm{\varepsilon}$ where $\Lambda$ is an $n \times n$ symmetric matrix \citep{rencher2008}.

We can express each of our quantities of interest as quadratic forms of $\bm{\varepsilon}$,
\begin{equation}
  \begin{aligned}
      \hat{C}_r &= \bm{\varepsilon}^T \Lambda_{\hat{C}_r} \bm{\varepsilon} \\
      \hat{C}_i &= \bm{\varepsilon}^T \Lambda_{\hat{C}_i} \bm{\varepsilon} \\
      \hat{P}_X &= \bm{\varepsilon}^T \Lambda_{\hat{P}_X} \bm{\varepsilon} \\
      \hat{P}_Y &= \bm{\varepsilon}^T \Lambda_{\hat{P}_Y} \bm{\varepsilon} \\
  \end{aligned}
\end{equation}
The explicit forms of these mixing matrices are obtained by substituting equations~(\ref{eq:X}) and (\ref{eq:Y}) into the definitions of the respective quantities ($\hat{C}_r$, $\hat{C}_i$, $\hat{P}_X$, and $\hat{P}_Y$) and expanding. We give the full matrix expressions in Appendix~\ref{app:mixing_matrices}.

We take advantage of the fact that quadratic forms of an $n$-dimensional normal variable with mean $\mu$ and covariance matrix $\Sigma_{\varepsilon}$ have expected value \citep{graybill1983,rencher2008}
\begin{equation}
    E[\bm{\varepsilon}^T \Lambda \bm{\varepsilon}] = \text{Tr}(\Lambda \Sigma_{\varepsilon}) + \mu^T \Lambda \mu,
\end{equation}
variance
\begin{equation}
    \text{Var}(\bm{\varepsilon}^T \Lambda \bm{\varepsilon})= 2 \text{Tr}(\Lambda \Sigma_{\varepsilon} \Lambda \Sigma_{\varepsilon}) + 4 \mu^T \Lambda \Sigma_{\varepsilon} \Lambda \mu,
\end{equation}
and covariance
\begin{equation}
    \text{Covar}(\bm{\varepsilon}^T \Lambda_1 \bm{\varepsilon}, \bm{\varepsilon}^T \Lambda_2 \bm{\varepsilon}) = 2 \text{Tr}(\Lambda_1 \Sigma_\varepsilon \Lambda_2 \Sigma_\varepsilon) + 4 \mu^T \Lambda_1 \Sigma_\varepsilon \Lambda_2 \mu,
\end{equation}
where Tr denotes the trace of a matrix.
Accordingly,
\begin{equation}
    \label{eq:expected_values}
    \begin{aligned}
        E[\hat{C}_r] &= \sqrt{P_X P_Y} \sqrt{\gamma^2} \cos \phi \\
        E[\hat{C}_i] &= \sqrt{P_X P_Y} \sqrt{\gamma^2} \sin \phi \\
        E[\hat{P}_X] &= P_X \\
        E[\hat{P}_Y] &= P_Y \\
    \end{aligned}
\end{equation}
and the covariance matrix in block form
\begin{equation}
\label{eq:covariance}
    \Sigma =
    \begin{pmatrix}
        \Sigma_{CC} & \Sigma_{CP} \\
        \Sigma_{CP}^T & \Sigma_{PP}
    \end{pmatrix}
\end{equation}
where the blocks are defined as follows: The cross-spectrum covariance block is
\begin{equation}
    \Sigma_{CC} =
    \begin{pmatrix}
        \frac{1}{2}P_X P_Y(1 + \gamma^2\cos 2\phi) & P_X P_Y\gamma^2\cos\phi\sin\phi \\
        P_X P_Y\gamma^2\cos\phi\sin\phi & \frac{1}{2}P_X P_Y(1 - \gamma^2\cos 2\phi)
    \end{pmatrix},
\end{equation}
the cross-power covariance block is
\begin{equation}
    \Sigma_{CP} =
    \begin{pmatrix}
        \sqrt{P_X^3 P_Y}\,\sqrt{\gamma^2}\cos\phi & \sqrt{P_X P_Y^3}\,\sqrt{\gamma^2}\cos\phi \\
        \sqrt{P_X^3 P_Y}\,\sqrt{\gamma^2}\sin\phi & \sqrt{P_X P_Y^3}\,\sqrt{\gamma^2}\sin\phi
    \end{pmatrix},
\end{equation}
and the power spectrum covariance block is
\begin{equation}
    \Sigma_{PP} =
    \begin{pmatrix}
        P_X^2 & P_X P_Y\gamma^2 \\
        P_X P_Y\gamma^2 & P_Y^2
    \end{pmatrix}.
\end{equation}
This covariance is valid for a single realization of $(\hat{C}_r, \hat{C}_i, \hat{P}_X, \hat{P}_Y)$. For $n$ realizations averaged together, $\Sigma_n = \dfrac{\Sigma}{n}$.

\subsection{Variance of the Coherence and Phase Lag}\label{sec:variance_equations}
Applying the delta method, the variance of the coherence and phase lag estimators is approximated by
\begin{equation}
    \begin{aligned}
        \text{Var}(\hat{\gamma}^2) &\approx (\nabla \coh)^T \cdot \frac{\Sigma}{n} \cdot \nabla \coh \\
        \text{Var}(\hat{\phi}) &\approx (\nabla \phi)^T \cdot \frac{\Sigma}{n} \cdot \nabla \phi
    \end{aligned}
\end{equation}
where gradients are taken with respect to $\bm{\hat{B}} = (\hat{C}_r, \hat{C}_i, \hat{P}_X, \hat{P}_Y)$ and evaluated at $\bm{B} = (\sqrt{P_X P_Y} \gamma \cos \phi, \sqrt{P_X P_Y} \gamma \sin \phi, P_X, P_Y)$. Evaluating these expressions, we find
\begin{equation}
    \label{eq:variances}
    \begin{aligned}
        \text{Var}(\cohhat) &\approx \frac{2 \coh (1-\coh)^2}{n}\\
        \text{Var}(\hat{\phi}) &\approx \frac{1-\coh}{2 n \coh}.\\
    \end{aligned}
\end{equation}
The variance of the coherence is in agreement with the result of \citet{bendat2010}.

\subsection{Covariance Structure of Coherence, Phase Lag, and Spectral Quantities}\label{sec:covariance_equations}

When jointly fitting models to power spectra and cross-spectra, the correlations between estimated quantities must be accounted for. We extend the delta method to compute the full covariance structure relating $\cohhat$ and $\hat{\phi}$ to the base quantities $(\hat{C}_r, \hat{C}_i, \hat{P}_X, \hat{P}_Y)$. The complete derivation is given in Appendix~\ref{app:covariance_structure}; we summarize the results here.

The coherence and phase lag estimators are uncorrelated to leading order in the delta method: $\text{Cov}(\cohhat, \hat{\phi}) \approx 0$ for all values of $\coh$ and $\phi$. This is consistent with the geometric intuition that coherence depends on the magnitude of the cross-spectrum while the phase lag depends on its argument.
The coherence estimator does exhibit weak correlations with all four base quantities, proportional to $\gamma^2(1-\gamma^2)$, which vanish at both zero and unit coherence. The phase lag estimator is uncorrelated with both power spectra.

The full $6 \times 6$ covariance matrix for the extended estimator vector $(\hat{C}_r, \hat{C}_i, \hat{P}_X, \hat{P}_Y, \cohhat, \hat{\phi})$ is given in equation~(\ref{eq:full_covariance}). When fitting models directly to the real and imaginary parts of the cross-spectrum along with the power spectra, as done, e.g., by \citet{mendez2024}, the off-diagonal terms in $\Sigma$ must be incorporated into the likelihood function. A diagonal likelihood which treats the power and cross spectra as independent is inadequate.
It should, however, be noted that the covariance matrix given in equation~(\ref{eq:full_covariance}) is a Gaussian approximation of the true covariance matrix. For a complete treatment and practical recommendations on implementing a likelihood function which accounts for these correlations, see \citet{nathan2026}.

\section{Results}
\label{sec:results}

\subsection{Application}

We apply the simulation method to a model with two Lorentzian components, each with a distinct phase lag. Both PSDs contain Lorentzian features centered at $\nu_0 = 1$\,Hz and $\nu_1 = 50$\,Hz with quality factors $Q_0 = 0.4$ and $Q_1 = 1.0$. The fractional normalization amplitudes are $A_0 = 0.012$ and $A_1 = 0.01$ in the reference band, and $B_0 = 0.05$ and $B_1 = 0.005$ in the dependent band. These parameters produce different PSD shapes in the two bands while yielding high coherence. The target coherence and phase lag profiles, which are the inputs of the simulation method, are given by the constant phase lag models of \citet{mendez2024}. These models describe the net phase lag at a given frequency as a sum of the phase lags of the Lorentzian components, weighted by their amplitudes at that frequency. When calculating the coherence profile, these models assume that individual Lorentzian components are coherent with themselves but mutually incoherent. This fact produces coherence drops at frequencies where multiple components contribute comparable power. We set component-specific phase lags to $\phi_0 = 0.15$\,rad and $\phi_1 = -0.8$\,rad.

For each realization, we simulate light curves with $N = 2^{18}$ time bins at resolution $\Delta t = 0.001$\,s (total duration $T \sim 262$\,s), mean count rate of $1000\,\mathrm{counts}\,\mathrm{s}^{-1}$, and target fractional RMS of 20\%. We compute averaged power spectra and cross-spectra using 16\,s segments (yielding $\sim$16 segments per realization) and apply logarithmic binning with a factor of 0.02. We generate $N_\mathrm{real} = 10$ independent realizations and compute ensemble-averaged estimates.

\begin{figure*}
    \includegraphics[width=0.8\textwidth]{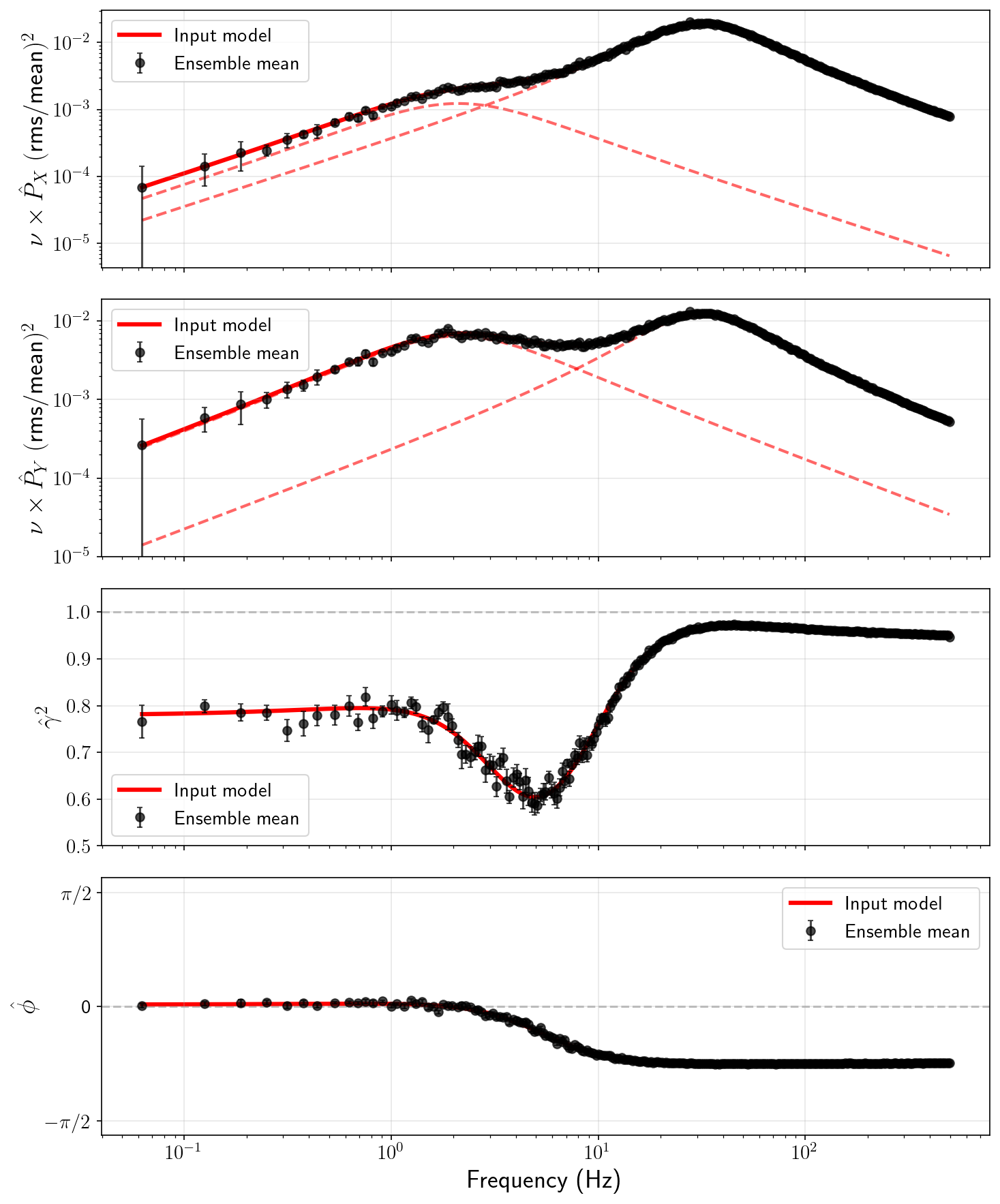}
    \caption{Ensemble validation of the simulation method applied to a realistic two-component model. Panels show (top to bottom): reference band PSD $\hat{P}_X$, dependent band PSD $\hat{P}_Y$, coherence $\hat{\gamma}^2$, and phase lag $\hat{\phi}$. Black points with error bars represent ensemble means $\pm$ standard errors over $N_\mathrm{real} = 10$ independent realizations. Solid red curves show scaled input models; dashed red curves in the PSD panels indicate individual Lorentzian components. The coherence profile follows the \citet{mendez2024} multi-component framework with component-specific phase lags $\phi_0 = 0.15$\,rad (1 Hz feature) and $\phi_1 = -0.8$\,rad (50\,Hz feature). Ensemble means track input models closely across all quantities, validating the simulation algorithm.}
    \label{fig:usage_example}
\end{figure*}

Fig.~\ref{fig:usage_example} shows the ensemble validation results. 
The top two panels display $\nu \times \hat{P}_X$ and $\nu \times \hat{P}_Y$. Black points with error bars represent ensemble means and standard errors. Solid red curves show the input models scaled to match the target RMS. Dashed red curves show the individual Lorentzian components. Both panels show close agreement between ensemble means and input models across $0.01$--$100$\,Hz, with Lorentzian features clearly visible near 1\,Hz and\,50 Hz.

The third panel shows $\hat{\gamma}^2(\nu)$. Where a single component dominates, the coherence approaches unity. At frequencies where both components contribute comparable power, the coherence drops due to their mutual incoherence. This pattern of high coherence except at component crossover regions has been observed in Cyg~X-1 and other black hole X-ray binaries \citep{nowak1999b,mendez2024,fogantini2025}.

The bottom panel shows $\hat{\phi}(\nu)$. The ensemble mean tracks the input model, with distinct lag features at the two frequencies: $\sim$0.15\,rad at 1\,Hz and ${\sim} -0.8$ rad at 50\,Hz. Where one component dominates, the phase lag reflects that component's intrinsic lag. The sharp transitions occur where power contributions shift between components.

\subsection{Monte Carlo Simulations}
We employ Monte Carlo simulations to verify that the variance formulas from Section~\ref{sec:variance} match observed scatter in simulated data and that coherence estimates exhibit the systematic bias from Section~\ref{sec:bias}.

For the variance tests, we generated 50\,000 realizations of single-frequency Fourier coefficients, each averaged over $n=50$ segments. We set the true coherence to $\gamma^2 = 0.75$ and phase lag to $\phi = \pi/4$, with power spectra $P_X = 1.0$ and $P_Y = 10.0$.
For each realization, we generate the random vector $\bm{\epsilon} \sim \mathcal{N}(\bm{0},\mathcal{I}_4)$ using the \texttt{numpy} implementation of the PCG-64 random number generator \citep{numpy}. We then compute the averaged power spectra ($\hat{P}_X$, $\hat{P}_Y$), cross-spectrum components ($\hat{C}_r$, $\hat{C}_i$), coherence $\cohhat$, and phase lag $\hat{\phi}$.

\begin{figure*}
    \centering
    \includegraphics[width=0.95\textwidth]{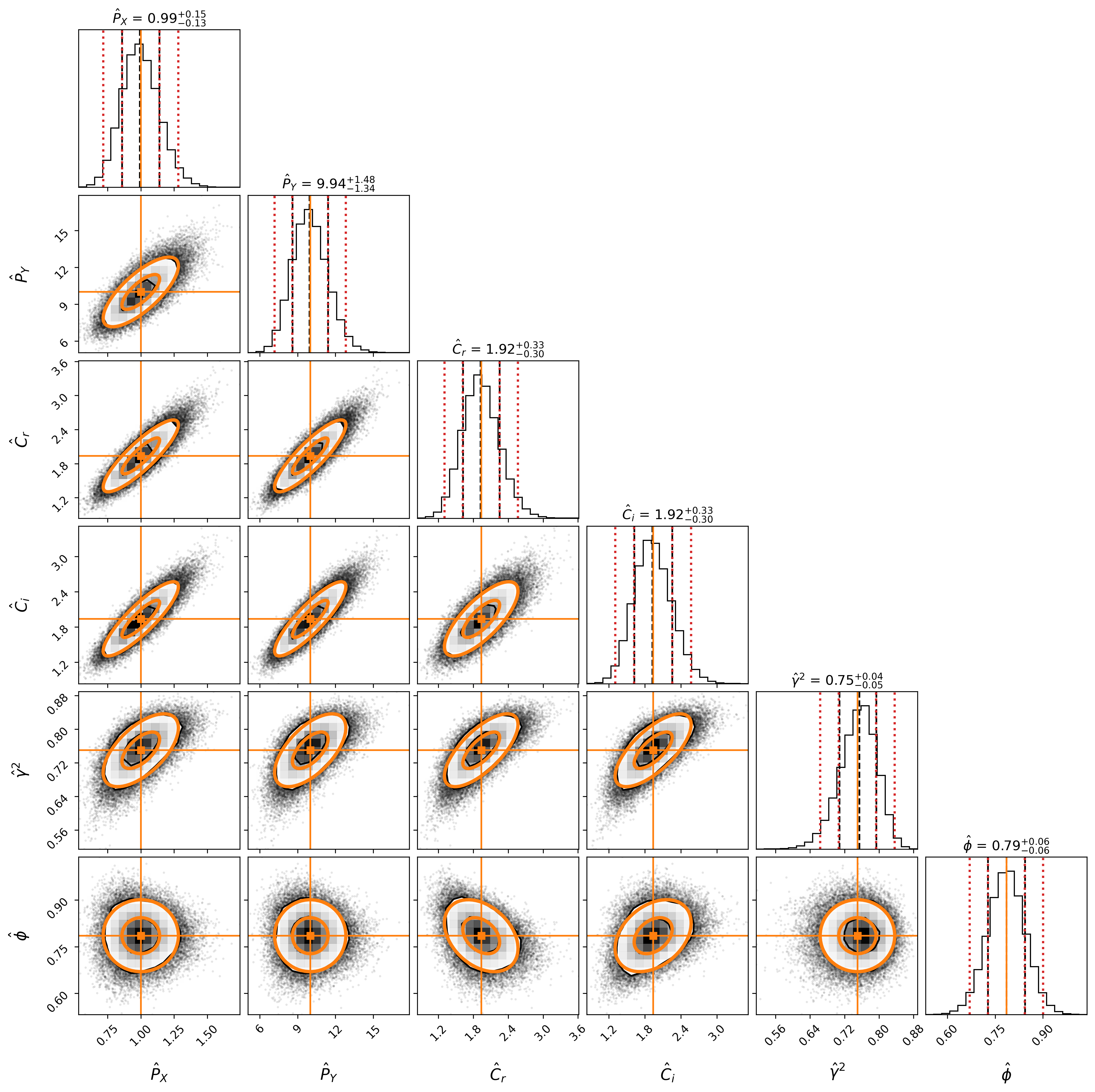}
    \caption{Joint distribution of the six estimators: base measurements ($\hat{P}_X$, $\hat{P}_Y$, $\hat{C}_r$, $\hat{C}_i$), and derived quantities ($\cohhat$, $\hat{\phi}$). Distributions generated from 50\,000 Monte Carlo realizations each averaging 50 instances of a single frequency. In the off diagonal panels, orange lines mark true values and ellipses show $1\sigma$ and $2\sigma$ contours from the full $6 \times 6$ covariance matrix (equation~\ref{eq:full_covariance}). Gray points each mark a single Monte Carlo realization and black contours mark the measured $1\sigma$ and $2\sigma$ levels. On-diagonal panels show marginalized distributions for each quantity with observed mean and standard deviation (dashed black lines). Solid orange lines mark the true value of each quantity (equation~\ref{eq:expected_values}) and dashed orange lines show theoretical standard deviations.}
    \label{fig:corner_all}
\end{figure*}

Fig.~\ref{fig:corner_all} shows the joint distributions of all six estimated quantities.
The marginal distributions are centered near the true values, with the coherence estimator being offset as predicted by the bias, equation~(\ref{eq:bias}).
The coherence estimator exhibits weak correlations with all four base quantities, as predicted by the covariance formulas in Appendix~\ref{app:covariance_structure}.
The phase lag estimator shows no correlation with the power spectra, depending only on the argument of the cross-spectrum rather than its normalization. While the predicted covariances match the simulations well, it can be observed that some quantities have non-Gaussian covariances. This non-Gaussian structure causes deviation from the predictions, which are Gaussian approximations.

\begin{figure*}
    \centering
    \includegraphics[width=0.85\textwidth]{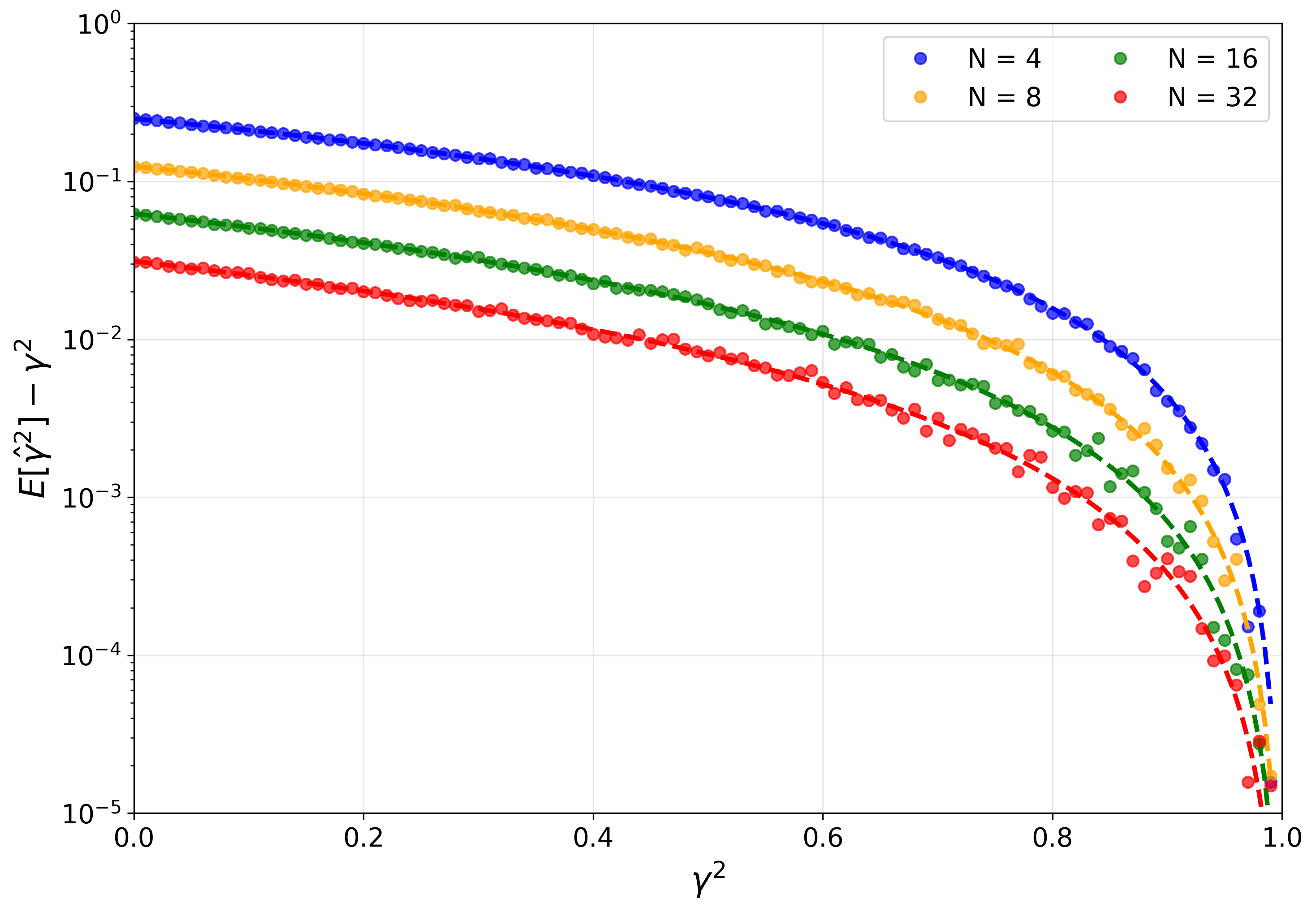}
    \caption{Bias $E[\cohhat]$--$\gamma^2$ versus true coherence for different numbers of averaged segments ($n = 4$, 8, 16, 32). Points show Monte Carlo results from 50\,000 trials at each coherence value ranging from 0 to 0.999; dashed lines show the exact theoretical bias from \citet{NuttallCarter1976}. The bias peaks at $\gamma^2=0$, where $E[\cohhat] = 1/n$, vanishes at $\gamma^2=1$, and scales approximately as $(1-\gamma^2)^2/n$ for intermediate coherence.}
    \label{fig:bias}
\end{figure*}

For the bias tests, we simulated coherence estimates across the full range of true coherence values for four different segment counts ($n = 4, 8, 16, 32$). Fig.~\ref{fig:bias} confirms that the simulation method reproduces the theoretical bias derived by \citet{NuttallCarter1976}. The practical implication is that low-$n$ measurements systematically overestimate coherence, particularly when the true coherence or number of segments averaged is low.

\section{Conclusions}
\label{sec:conclusions}

We have presented a method for generating pairs of synthetic time series with prescribed power spectra, coherence, and phase lag profiles. The approach generalizes the \citet{timmer1995} algorithm by decomposing the dependent time series into coherent and incoherent components. The coherent component is constructed via a complex transfer function applied to the reference Fourier transform, which is then normalized to ensure the target power spectrum is preserved. This construction allows independent specification of $P_X(\nu)$, $P_Y(\nu)$, $\gamma^2(\nu)$, and $\phi(\nu)$.

We derived closed-form expressions for the variances of the coherence and phase lag estimators (equation~\ref{eq:variances}). We extended this analysis to the full $6 \times 6$ covariance matrix relating the base quantities $(\hat{C}_r, \hat{C}_i, \hat{P}_X, \hat{P}_Y)$ and derived quantities $(\cohhat, \hat{\phi})$ in equation~(\ref{eq:full_covariance}). Monte Carlo simulations confirm these predictions. The phase lag estimator is uncorrelated with both power spectra, depending only on the argument of the cross-spectrum. The coherence estimator exhibits weak correlations with all four base quantities at intermediate $\gamma^2$, vanishing at both zero and unit coherence.

The covariance structure has direct implications for cross-spectral fitting. When jointly modeling $C_r$, $C_i$, $P_X$, and $P_Y$ as done, e.g., by \citet{mendez2024}, the off-diagonal terms in $\Sigma$ (equation~\ref{eq:covariance}) must be incorporated into the likelihood function. 

The method as presented has limitations. It generates noiseless time series; for realistic XRB or AGN simulations, Poisson counting noise must be added after generating the underlying signal. Poisson noise adds a flat (white) power component to each band's PSD at a level set by the mean count rate. Since the noise is uncorrelated between bands, the cross-spectrum is unchanged and the measured phase lag equals its intrinsic value. The coherence is diluted relative to the intrinsic value as
\begin{equation}
    \gamma^2_{\rm obs}(\nu) = \gamma^2_{\rm int}(\nu) \cdot \frac{P^s_X(\nu)}{P^s_X(\nu) + P_{\rm noise,X}} \cdot \frac{P^s_Y(\nu)}{P^s_Y(\nu) + P_{\rm noise,Y}},
    \label{eq:poisson_dilution}
\end{equation}
where $P^s_{X/Y}(\nu)$ are the noiseless signal power spectra and $P_{\rm noise, X/Y}$ are the Poisson noise floors \citep[e.g.,][]{vaughan1997,nowak1999a}. 
Beyond noise, the detector response and time-binning process can themselves introduce apparent coherence between intrinsically uncorrelated signals, as discussed in Appendix~\ref{app:detector_coherence}. Further detector effects, such as pile-up \citep{davis2001,sevilla2017} and dead time \citep{zhang1995,huppenkothen2022} alter observed time series and Fourier products in ways not accounted for in raw simulation products. For these reasons, a stage of processing to adjust simulation outputs to better resemble realistic data -- adding Poisson noise and other detector effects -- is recommended. 

If the generated time series are segmented and averaged before being analyzed, the effects of spectral leakage must be contended with, as discussed in Appendix~\ref{app:spectral-leakage}. Additionally, the construction also captures only linear correlations between bands, neglecting any higher-order terms.

The delta method variance expressions are large-$n$ approximations. At small $n$, the coherence estimator becomes non-Gaussian and the bias from Section~\ref{sec:bias} becomes substantial. The \citet{NuttallCarter1976} bias correction can be applied post hoc by calculating this bias and subtracting it from the coherence function of the simulated data. Finally, the method produces Gaussian time series and does not reproduce the log-normal flux distributions observed in accreting systems \citep{uttley2005}. Extending to non-Gaussian statistics would require additional transformations, such as those developed by \citet{emmanoulopoulos2013}.

Finally, we stress that the derived transfer function $R$ (equation \ref{eq:R}) is a mathematical transfer function relating the two time series and does not, in general, correspond to a physical response function. When \textbf{y} leads \textbf{x} ($\phi > 0$), $R$ encodes a causality-violating relationship between the time series where the dependent ``responds'' to the future of the reference. For this reason, we caution against interpreting $R$ as encoding physical information about the system. This construction is therefore appropriate for generating synthetic time series which reproduce target observables, not for imposing a physical model on the signal generation process. 

The method enables several applications. Synthetic data can validate lag-recovery pipelines used in reverberation mapping studies and test multi-Lorentzian cross-spectral fitting frameworks under controlled conditions. The variance formulas provide \textit{a priori} estimates of how many segments are needed to achieve target precision on $\gamma^2$ or $\phi$, informing experiment design.

Cross-spectral techniques are central to interpreting variability in XRBs and AGN. Recent discoveries of hidden QPO-like features
with large phase lags \citep{mendez2024,konig2024,fogantini2025} underscore the diagnostic power of coherence and phase lag measurements. As these techniques become more sophisticated, robust simulation tools with well-characterized statistical properties become necessary. The method presented here provides a foundation for these techniques.

\section*{Conflicts of Interest}
The authors declare no conflicts of interest.

\section*{Acknowledgments}
\edit{We thank Martin Luepker for his insightful comments.}
SRL and MAN acknowledge funding from NASA grant 80NSSC24K0616. \edit{We acknowledge the anonymous reviewers whose comments have improved this paper.}

\section*{Data Availability}

A code implementing the method of this paper, designed to work with the open-source X-ray timing package \texttt{stingray} \edit{\citep{stingray1,stingray2,stingraysoftware}}, is available on GitHub at \texttt{https://github.com/SethHtes/synthetic-timeseries}. The scripts to produce the figures in this paper, along with the Wolfram Mathematica \edit{\citep{Mathematica}} notebooks used in the derivation, can be found at \texttt{https://github.com/SethHtes/RASTI-scripts}.



\bibliographystyle{rasti}
\bibliography{bib} 



\appendix

\section{Mixing Matrices}
\label{app:mixing_matrices}

The covariance matrix $\Sigma$ for the vector of estimators $\bm{\hat{B}} = (\hat{C}_r, \hat{C}_i, \hat{P}_X, \hat{P}_Y)$ is derived by expressing each quantity as a quadratic form of the random variable $\bm{\varepsilon} = (A_r, B_r, H_r, J_r)$. Specifically, each estimator can be written as $\bm{\varepsilon}^T \Lambda \bm{\varepsilon}$, where $\Lambda$ is a symmetric mixing matrix.

These matrices are constructed by substituting the Fourier transform expressions from equations~(\ref{eq:X}) and (\ref{eq:Y}) into the definitions of the cross-spectrum components and power spectra. For the real part of the cross-spectrum, $\hat{C}_r = \text{Re}[X^* Y]$; for the imaginary part, $\hat{C}_i = \text{Im}[X^* Y]$; and for the power spectra, $\hat{P}_X = X X^*$ and $\hat{P}_Y = Y Y^*$. Expanding these products in terms of the random variables $(A_r, B_r, H_r, J_r)$ and rearranging into quadratic form yields the mixing matrices below.

For compact notation, define $\alpha \equiv \sqrt{\gamma^2}$ and $\beta \equiv \sqrt{1-\gamma^2}$.
The mixing matrix for the real part of the cross-spectrum is
\begin{equation}
    \Lambda_{\hat{C}_r} = \frac{\sqrt{P_X P_Y}}{4}
    \begin{pmatrix}
        2\alpha\cos\phi & 0 & \beta & 0 \\
        0 & 2\alpha\cos\phi & 0 & \beta \\
        \beta & 0 & 0 & 0 \\
        0 & \beta & 0 & 0
    \end{pmatrix}.
\end{equation}
The mixing matrix for the imaginary part of the cross-spectrum is
\begin{equation}
    \Lambda_{\hat{C}_i} = \frac{\sqrt{P_X P_Y}}{4}
    \begin{pmatrix}
        2\alpha\sin\phi & 0 & 0 & \beta \\
        0 & 2\alpha\sin\phi & -\beta & 0 \\
        0 & -\beta & 0 & 0 \\
        \beta & 0 & 0 & 0
    \end{pmatrix}.
\end{equation}
The mixing matrix for the reference power spectrum is
\begin{equation}
    \Lambda_{\hat{P}_X} = \frac{P_X}{2}
    \begin{pmatrix}
        1 & 0 & 0 & 0 \\
        0 & 1 & 0 & 0 \\
        0 & 0 & 0 & 0 \\
        0 & 0 & 0 & 0
    \end{pmatrix}.
\end{equation}
The mixing matrix for the dependent power spectrum is
\begin{equation}
    \Lambda_{\hat{P}_Y} = \frac{P_Y}{2}
    \begin{pmatrix}
        \alpha^2 & 0 & \alpha\beta\cos\phi & \alpha\beta\sin\phi \\
        0 & \alpha^2 & -\alpha\beta\sin\phi & \alpha\beta\cos\phi \\
        \alpha\beta\cos\phi & -\alpha\beta\sin\phi & \beta^2 & 0 \\
        \alpha\beta\sin\phi & \alpha\beta\cos\phi & 0 & \beta^2
    \end{pmatrix}
\end{equation}
where $\alpha = \sqrt{\gamma^2}$ and $\beta = \sqrt{1-\gamma^2}$ as defined above.

Using the properties of quadratic forms of multivariate normal random variables \citep{graybill1983,rencher2008}, the covariance matrix $\Sigma$ in equation~(\ref{eq:covariance}) can then be computed from these mixing matrices.

\section{Covariance Structure of Coherence, Phase Lag, and Spectral Quantities}
\label{app:covariance_structure}

We extend the delta method (Section~\ref{sec:variance}) to compute the covariances between the derived quantities $\cohhat$ and $\hat{\phi}$ and the base quantities $(\hat{C}_r, \hat{C}_i, \hat{P}_X, \hat{P}_Y)$. For two functions $f(\bm{\hat{B}})$ and $g(\bm{\hat{B}})$ of the estimator vector, the covariance is
\begin{equation}
    \text{Cov}(f(\bm{\hat{B}}), g(\bm{\hat{B}})) \approx (\nabla f)^T \cdot \frac{\Sigma}{n} \cdot \nabla g
\end{equation}
where gradients are evaluated at the true value, $\boldsymbol{B}$ \citep{graybill1983}.

The gradients of the coherence and phase lag with respect to the base quantities are
\begin{equation}
    \begin{aligned}
        \nabla \coh &= \left( \frac{2\gamma\cos\phi}{\sqrt{P_X P_Y}}, \frac{2\gamma\sin\phi}{\sqrt{P_X P_Y}}, -\frac{\gamma^2}{P_X}, -\frac{\gamma^2}{P_Y} \right) \\
        \nabla \phi &= \left( -\frac{\sin\phi}{\gamma\sqrt{P_X P_Y}}, \frac{\cos\phi}{\gamma\sqrt{P_X P_Y}}, 0, 0 \right).
    \end{aligned}
\end{equation}

\subsection{Covariance Between Coherence and Phase Lag}

Evaluating the covariance between coherence and phase lag yields
\begin{equation}
    \text{Cov}(\cohhat, \hat{\phi}) \approx (\nabla \coh)^T \cdot \frac{\Sigma}{n} \cdot \nabla \phi = 0.
\end{equation}
The two estimators are therefore uncorrelated to leading order in the delta method for all values of $\coh$ and $\phi$. This is consistent with the geometric intuition that coherence depends on the magnitude of the cross-spectrum while phase lag depends on its argument.

\subsection{Covariances with Spectral Quantities}

The covariances between the coherence estimator and the underlying spectral quantities are
\begin{equation}
    \label{eq:coh_base_cov}
    \begin{aligned}
        \text{Cov}(\cohhat, \hat{C}_r) &\approx \frac{\sqrt{P_X P_Y}\,\sqrt{\gamma^2}(1 - \gamma^2)\cos\phi}{n} \\
        \text{Cov}(\cohhat, \hat{C}_i) &\approx \frac{\sqrt{P_X P_Y}\,\sqrt{\gamma^2}(1 - \gamma^2)\sin\phi}{n} \\
        \text{Cov}(\cohhat, \hat{P}_X) &\approx \frac{P_X \gamma^2 (1 - \gamma^2)}{n} \\
        \text{Cov}(\cohhat, \hat{P}_Y) &\approx \frac{P_Y \gamma^2 (1 - \gamma^2)}{n}.
    \end{aligned}
\end{equation}
To leading order in the delta method, all covariances with the coherence vanish in the limits $\gamma^2 \to 0$ and $\gamma^2 \to 1$, and reach a maximum at intermediate coherence.

For the phase lag estimator,
\begin{equation}
    \label{eq:phi_base_cov}
    \begin{aligned}
        \text{Cov}(\hat{\phi}, \hat{C}_r) &\approx -\frac{\sqrt{P_X P_Y}(1 - \gamma^2)\sin\phi}{2n\sqrt{\gamma^2}} \\
        \text{Cov}(\hat{\phi}, \hat{C}_i) &\approx \frac{\sqrt{P_X P_Y}(1 - \gamma^2)\cos\phi}{2n\sqrt{\gamma^2}} \\
        \text{Cov}(\hat{\phi}, \hat{P}_X) &\approx 0 \\
        \text{Cov}(\hat{\phi}, \hat{P}_Y) &\approx 0.
    \end{aligned}
\end{equation}
To leading order in the delta method, the phase lag is uncorrelated with both power spectra, consistent with the geometric intuition that the phase depends only on the argument of the cross-spectrum, not on the magnitude.

\subsection{Full Covariance Matrix}

Combining these results with the covariance matrix $\Sigma$ from equation~(\ref{eq:covariance}), the full $6 \times 6$ covariance matrix for the extended estimator vector $(\hat{C}_r, \hat{C}_i, \hat{P}_X, \hat{P}_Y, \cohhat, \hat{\phi})$ can be written in block form as
\begin{equation}
\label{eq:full_covariance}
    \Sigma_{\text{full}} =
    \begin{pmatrix}
        \Sigma & \Sigma_{B\gamma} & \Sigma_{B\phi} \\
        \Sigma_{B\gamma}^T & \Sigma_{\gamma\gamma} & 0 \\
        \Sigma_{B\phi}^T & 0 & \Sigma_{\phi\phi}
    \end{pmatrix}
\end{equation}
where $\Sigma$ is the $4 \times 4$ base covariance matrix from equation~(\ref{eq:covariance}), and the additional blocks are defined as follows. The coherence-base covariance block is the $4 \times 1$ column vector
\begin{equation}
    \Sigma_{B\gamma} =
    \begin{pmatrix}
        \text{Cov}(\cohhat, \hat{C}_r) \\
        \text{Cov}(\cohhat, \hat{C}_i) \\
        \text{Cov}(\cohhat, \hat{P}_X) \\
        \text{Cov}(\cohhat, \hat{P}_Y)
    \end{pmatrix},
\end{equation}
the phase lag-base covariance block is the $4 \times 1$ column vector
\begin{equation}
    \Sigma_{B\phi} =
    \begin{pmatrix}
        \text{Cov}(\hat{\phi}, \hat{C}_r) \\
        \text{Cov}(\hat{\phi}, \hat{C}_i) \\
        \text{Cov}(\hat{\phi}, \hat{P}_X) \\
        \text{Cov}(\hat{\phi}, \hat{P}_Y)
    \end{pmatrix},
\end{equation}
and the derived quantity variances are the scalars $\Sigma_{\gamma\gamma} = \text{Var}(\cohhat)$ and $\Sigma_{\phi\phi} = \text{Var}(\hat{\phi})$. The derived quantities $\cohhat$ and $\hat{\phi}$ inherit correlated uncertainties with the base quantities, even though they are uncorrelated with each other to leading order in the delta method.

\section{Effects of Detection and Discretization on Coherence}
\label{app:detector_coherence}

The coherence function defined in equation~(\ref{eq:coherence}) characterizes the linear correlation between two time series at each Fourier frequency. Throughout the main text, we work with the intrinsic coherence of the underlying astrophysical signals, prior to any measurement process. In practice, the signals we observe have been filtered through the detector response and binned into discrete time samples. Both of these steps can alter the measured coherence relative to its intrinsic value, and in some cases they can introduce apparent coherence between signals that are intrinsically uncorrelated.

\subsection{Detector Response}

Consider two energy bands measured by the same detector. A photon arriving at the detector undergoes energy redistribution, described by the response matrix $\mathcal{R}(E', E)$ giving the probability that a photon of true energy $E$ is recorded at measured energy $E'$ \citep{arnaud1996}.
We focus here on photon energy redistribution; additional instrumental effects such as pile-up and dead time can introduce further correlations between energy bands but are beyond the scope of this treatment.

Suppose the source emission consists of two physically independent components with Fourier transforms $S_1(\nu)$ and $S_2(\nu)$, where $E[S_1^* S_2] = 0$. After passing through the detector, the signal recorded in energy band $X$ is a weighted sum of contributions from both components,
\begin{equation}
    \begin{aligned}
    X(\nu) &= \alpha_{X1}\, S_1(\nu) + \alpha_{X2}\, S_2(\nu) \\
    Y(\nu) &= \alpha_{Y1}\, S_1(\nu) + \alpha_{Y2}\, S_2(\nu),
    \end{aligned}
\end{equation}
where the coefficients $\alpha$ are determined by the detector response integrated over the respective energy bands. The cross-spectrum between the two measured bands is then
\begin{equation}
\label{eq:detector_cross}
    C(\nu) = E[X^* Y] = \alpha_{X1}^* \alpha_{Y1}\, P_{S_1}(\nu) + \alpha_{X2}^* \alpha_{Y2}\, P_{S_2}(\nu).
\end{equation}
Although $S_1$ and $S_2$ are mutually uncorrelated, the cross-spectrum and hence the coherence between bands $X$ and $Y$ is generally nonzero, because both bands receive contributions from both source components. The measured coherence reflects the way the detector mixes source components into the observed energy channels, not only the intrinsic correlation structure of the source.

This is most transparent in an extreme case. Consider two uncorrelated sources, each producing a single impulsive event (a delta function in time) at the same instant $t_0$. Both events arrive simultaneously and pass through the same detector, so the recorded signals in each band are
\begin{equation}
    \begin{aligned}
        x(t) &= (\alpha_{X1} + \alpha_{X2})\, \delta(t - t_0) \\
        y(t) &= (\alpha_{Y1} + \alpha_{Y2})\, \delta(t - t_0).
    \end{aligned}
\end{equation}
The measured signals are proportional to each other at all times, so their coherence is identically unity. The two sources are intrinsically unrelated, yet the measurement process makes them indistinguishable from a single perfectly coherent source. No amount of averaging can separate them. This deterministic example is degenerate but serves to illustrate the mechanism transparently. The general effect operates for continuous stochastic processes as described in Section C.1: once the detector mixes both components into both bands, the stochastic statement $E[S_1^* S_2] = 0$ becomes irrelevant because the mixing creates a deterministic linear relationship between the measured signals.

Any time two independent signals are detected by the same instrument and contribute to the same observed energy bands, the shared measurement process creates correlations in the output that are absent in the input.

\subsection{Discretization and Aliasing}

Constructing a light curve from photon event data involves two operations: integrating the arrival rate over time bins of width $\Delta t$, and sampling the result at discrete intervals. The bin-integration convolves the continuous signal with a rectangular window, which in the Fourier domain multiplies all components by $W(\nu) = \Delta t\, \mathrm{sinc}(\pi \nu \Delta t)$, where $\mathrm{sinc}(x) \equiv \sin(x)/x$. This multiplicative factor is common to everything passing through the same binning process and cancels in the coherence ratio, assuming both bands are binned identically. The bin-integration does suppress high-frequency power through the $\mathrm{sinc}^2$ envelope, but it does not, on its own, alter the measured coherence at any frequency.

The discrete sampling does modify the coherence. Sampling at rate $\nu_s = 1/\Delta t$ folds power from every frequency $\nu + k\nu_s$ ($k$ integer) onto the measured frequency $\nu$. The power spectrum of the discretized time series at a frequency $0 < \nu < \nu_{\mathrm{Nyq}} = \nu_s/2$ is not simply the continuous PSD evaluated at $\nu$, but the aliased sum \citep[e.g.,][]{percival1993,bendat2010}
\begin{equation}
\label{eq:aliased_psd}
    P_X^\mathrm{disc}(\nu) = \sum_{k=-\infty}^{\infty} w_k(\nu) \, P_X^\mathrm{cont}(\nu + k\nu_s)
\end{equation}
where $w_k(\nu) \equiv |W(\nu + k\nu_s)|^2$ are the aliasing weights and $P_X^\mathrm{cont}$ is the continuous (pre-sampling) PSD. The cross-spectrum obeys the same relation,
\begin{equation}
\label{eq:aliased_csd}
    C^\mathrm{disc}(\nu) = \sum_{k=-\infty}^{\infty} w_k(\nu) \, C^\mathrm{cont}(\nu + k\nu_s).
\end{equation}
The weights decay with $|k|$ due to the $\mathrm{sinc}^2$ envelope but do not vanish, so every harmonic of $\nu_s$ contributes in principle.

The coherence of the discretized signals is then
\begin{equation}
\label{eq:disc_coherence}
    \gamma^2_\mathrm{disc}(\nu) = \frac{\left|\sum_k w_k \, C^\mathrm{cont}(\nu + k\nu_s)\right|^2}{\left(\sum_k w_k \, P_X^\mathrm{cont}(\nu + k\nu_s)\right) \left(\sum_k w_k \, P_Y^\mathrm{cont}(\nu + k\nu_s)\right)}.
\end{equation}
In general, $\gamma^2_\mathrm{disc}(\nu) \neq \gamma^2_\mathrm{cont}(\nu)$. Equality holds when $C^\mathrm{cont}$, $P_X^\mathrm{cont}$, and $P_Y^\mathrm{cont}$ share the same frequency dependence, which requires constant coherence, constant phase lag, and constant PSD ratio $P_Y/P_X$ across all aliased frequencies.

At each measured frequency, $\nu$, the discrete quantities receive contributions from both the base frequency ($k=0$) and the aliased harmonics ($k \neq 0$).
If the cross-spectral properties at the aliased frequencies differ from those at $\nu$, the aliased sum mixes these together and the resulting coherence reflects a weighted average rather than the value at $\nu$ alone. However, this effect can be diminished by choosing a high sampling rate \citep[see e.g., the discussion in][]{vanderklis1989}.

\subsection{Implications for Simulations}

The measured coherence is the coherence of the observed signals as processed by a particular instrument and set of analysis choices. It is not an absolute property of the astrophysical source. This distinction matters when multiple physically distinct emission components contribute to the observed bands, when the detector response redistributes photons across energy channels, or when the time binning is not much finer than the shortest variability timescale of interest.

The simulation method presented in this paper generates time series with a specified intrinsic coherence, before any instrumental effects are applied. To produce realistic simulated observations, one should convolve the output with the appropriate detector response and rebin to the desired time resolution before computing cross-spectral quantities. Further complicating effects may be captured using detector simulation methods \citep[e.g., SIXTE][]{dauser2019}. The resulting measured coherence will then generally differ from the input intrinsic coherence. 


\begin{edit-sec}
\section{Effects of Spectral Leakage due to Segmenting}
\label{app:spectral-leakage}

The algorithm derived in Section~\ref{sec:derivations} produces Fourier coefficients
that satisfy the target cross-spectral properties at every frequency bin. In practice however, time series are often analyzed in segments \citep[see e.g., the ``practical steps'' for analysis outlined in][]{uttley2014}. Long light curves are divided into equal length segments assuming stationarity and Fourier transformed. Average power- and cross-spectral quantities can then be calculated by averaging the properties of these segments together. This is of particular use when calculating the coherence function, which is trivially unity when applied to non-averaged power and cross-spectra. However, segmenting the full light curve introduces spectral leakage effects which manifest in recovered coherence and lag profiles when combined with the enforcement of Hermitian symmetry. We account for these effects in the context of the described simulation method, focusing on the effects on the coherence and phase lag profiles derived from algorithm outputs.

\subsection{The Dirichlet Kernel}
By the convolution theorem, segmenting the full time series is equivalent to convolving its Fourier transform with the Dirichlet kernel, the Fourier transform of a Dirac comb function. 
Consider a discrete time series $x_n$ of length $N$ bins with discrete Fourier transform $X_k$. Segment the time series into blocks of lengths $N_s$ to form segments $\tilde{x}_{n'}$ with Fourier transform $\tilde{X}_{k'}$. 
Throughout this Appendix, we distinguish segmented quantities from their full time series counterparts with $\tilde{\cdot}$ and by marking indices of segmented quantities with $\cdot'$.

The discrete Fourier transform of the segment is related to the full discrete Fourier transform through the convolution
\begin{equation}
    \begin{aligned}
    \tilde{X}_{j'} &= \frac{1}{N} \sum_{k=0}^{N-1} X_k \exp \left [ 2 \pi i k m / N\right ] W_{k,j'}\\
    &= \frac{1}{N} \sum_{k=0}^{N-1} X_k \alpha_k W_{k,j'}, 
    \end{aligned}
\end{equation}
where $m$ is the index of the start of the segment in $x_n$.
Then, the Dirichlet kernel describes the contribution of power from frequency bin $k$ in the unsegmented Fourier transform to frequency bin $j'$ in the segmented Fourier transform: 
\begin{equation}
    W_{k,j'} = e^{i\pi(N_s - 1) \left (\frac{k}{N}-\frac{j'}{N_s} \right )} \frac{\sin \left [N_s \pi \left (\frac{k}{N}-\frac{j'}{N_s} \right ) \right ]}{\sin \left [ \pi \left (\frac{k}{N}-\frac{j'}{N_s} \right )\right ]}.
\end{equation}

\subsection{Hermitian Symmetry}
\label{app:hermitian-symmetry}
By properties of the discrete Fourier transform, a real valued time series produces Hermitian-symmetric Fourier coefficients. For $x_n \in \mathds{R}, n = 0, \ldots, N-1$ we require $X_{k} = X^*_{N-k}$. For this reason, the method specified in this paper should be used to generate $N/2 - 1$ distinct complex Fourier coefficients for both the reference and dependent time series. By convention, we choose to describe these components as $X_k, k = 1, \ldots, N/2 -1$. When applying this method to simulating real time series, the remaining coefficients can be filled in according to the symmetry criterion. 

Components $k=0$ and $k = N/2$ are real by enforcement of Hermitian symmetry. $X_0$, known as the DC signal component, describes the overall level of signal in the time series
\begin{equation}
    X_0 = \sum_{n = 0}^{N-1} x_n,
\end{equation}
while $X_{N/2}$, the Nyquist frequency component, describes the highest measurable frequency. 

\subsection{Derivation of Leakage Effects}
The effects of the application of the Dirichlet kernel and the enforcement of Hermitian symmetry combine to effect simulated phase lag and coherence profiles when there is a non-zero phase lag between the dependent and reference time series. For the following derivation, we consider the case of simulating time series with constant phase lag profile $\phi(\nu_k) = \phi$ and unity coherence $\coh = 1$. 
We note that the effects discussed in this Appendix are applicable to general coherence and phase lag profiles, however detailed discussion of the more general case is beyond the scope of this Appendix. 
Finally, we will neglect the DC and Nyquist frequency components because $W_{0,j'}$ and $W_{N/2,j'}$ vanish for $j' \in [1,N_s/2-1]$.


By using Equations~\ref{eq:X}~and~\ref{eq:Y}, we can write 
\begin{equation}
    \begin{aligned}
        X_k &= 
        \begin{cases}
            \left ( \frac{P_{X,k}}{2} \right )^{1/2} (A_{r,k} + i B_{r,k}) & k \in [1,N/2-1] \\
            X^{*}_{N-k} &k \in [N/2+1,N-1]\\
        \end{cases} \\
        Y_k &= 
        \begin{cases}
            R_k X_k  & k \in [1,N/2-1] \\
            Y_{N-k}^* &k \in [N/2+1,N-1]\\
        \end{cases} \\
    \end{aligned}
\end{equation}
choosing to make explicit the implicit Hermitian symmetry convention. $R_k$ is given by Equations \ref{eq:R}. $A_r, B_r, H_r, J_r$ are independent and identically distributed (i.i.d.) standard normal variables.
We now segment each Fourier transform by applying the Dirichlet Kernel,

\begin{IEEEeqnarray}{rCl}
  \tilde{X}_{j'} &=& \frac{1}{N} \sum_{k=1}^{N-1} X_k \alpha_k W_{k,j'} \\
  &=& \frac{1}{N}\Biggl(\sum_{k=1}^{N/2-1} X_k \alpha_k W_{k,j'} \nonumber\\
  && \quad{}+ \sum_{k=N/2+1}^{N-1} X^*_{N-k} \alpha_k W_{k,j'}\Biggr) \nonumber\\
  &=& \frac{1}{N}\Biggl(\sum_{k=1}^{N/2-1} X_k \alpha_k W_{k,j'} \nonumber\\
  && \quad{}+ \sum_{l=1}^{N/2-1} X^*_l\, \alpha_{N-l} W_{N-l,j'}\Biggr) \nonumber
\end{IEEEeqnarray}
where we have re-indexed by $l = N - k$. We can similarly write the dependent Fourier transform,

\begin{IEEEeqnarray}{rCl}
  \tilde{Y}_{j'} &=& \frac{1}{N}\sum_{k=1}^{N-1} Y_k\,e^{2\pi i k m/N}\,W_{k,j'} \\
  &=& \frac{1}{N}\Biggl(e^{i\phi}\sum_{k=1}^{N/2-1} X_k\,\alpha_k\,W_{k,j'} \nonumber \\
  && \quad{}+ e^{-i\phi}\sum_{l=1}^{N/2-1} X_l^{*}\,\alpha_{N-l}\,W_{N-l,j'}\Biggr) \nonumber \\
\end{IEEEeqnarray}


Wishing to understand the behavior of the coherence and phase lag profiles, we turn now to the averaged segmented cross spectrum. Again approximating finite averaging with taking expectation values over random variables, we can write
\begin{IEEEeqnarray}{rcl}
    \langle \hat{\tilde{C}}_{j'}\rangle &\approx& E[\tilde{C}_{j}]\\
    &=& E[\tilde{X}^*_{j'} \tilde{Y}_{j'}] \nonumber \\
    &=& \frac{1}{N^2} E\Biggl[
        e^{i\phi} \sum_{k_1,k_2=1}^{N/2-1} X^*_{k_1} X_{k_2}\,
            \alpha_{k2-k1} W^*_{k_1,j'} W_{k_2,j'} \nonumber \\
    && \quad{} + e^{-i\phi} \sum_{k_1,l_2=1}^{N/2-1} X^*_{k_1} X^*_{l_2}\,
            \alpha_{k_1 + l_2} W^*_{k_1,j'} W_{N-l_2,j'} \nonumber \\
    && \quad{} + e^{i\phi} \sum_{l_1,k_2=1}^{N/2-1} X_{l_1} X_{k_2}\,
             \alpha_{k2+l1} W^*_{N-l_1,j'} W_{k_2,j'} \nonumber \\
    && \quad{} + e^{-i\phi} \sum_{l_1,l_2=1}^{N/2-1} X_{l_1} X^*_{l_2}\,
            \alpha_{l1-l2}W^*_{N-l_1,j'} W_{N-l_2,j'} \Biggr] \nonumber\\
    &=& \frac{1}{N^2}
        e^{i\phi} \sum_{k=1}^{N/2-1} P_{X,k} | W_{k,j'}|^2 + \frac{1}{N^2} e^{-i\phi} \sum_{l=1}^{N/2-1} P_{X,l} |W_{N-l,j'} |^2 \nonumber \\
    &=&  e^{i \phi} p_{j'} + e^{-i \phi} q_{j'}  \nonumber
\end{IEEEeqnarray}
where we have used the following relationships, readily derived from the properties of i.i.d complex normal variables: $E[X_{i}X_{j}] = E[X^*_{i}X^*_{j}] = 0$ and $E[X_{i}X^*_{j}] = P_X \delta_{i,j}$. For convenience, we define $p$ and $q$ to be power-weighted sums of Dirichlet kernels. 
A similar calculation yields 
\begin{equation}
    \langle \hat{\tilde{P}}_{X,j'}\rangle = \langle \hat{\tilde{P}}_{Y,j'}\rangle = p_{j'} + q_{j'}.
\end{equation}
Using these quantities, we can immediately write the forms of the estimated coherence and phase lag profiles:
\begin{equation} \label{eq:leaky_coh}
\begin{aligned}
    \hat{\tilde{\gamma}}^2_{j'} &= \frac{p^2_{j'} + q^2_{j'} + 2 p_{j'} q_{j'}\cos 2 \phi}{(p_{j'}+q_{j'})^2}\\
    &= \frac{1 + r_{j'}^2 + 2r_{j'} \cos 2 \phi}{(1 + r_{j'})^2}
\end{aligned}
\end{equation}

\begin{equation} \label{eq:leaky_phi}
\begin{aligned}
    \hat{\tilde{\phi}}_{j'} &= \arctan \left [ \frac{(p_{j'} - q_{j'}) \sin \phi}{(p_{j'} + q_{j'}) \cos \phi}\right]\\
    &= \arctan \left [ \frac{(1 - r_{j'}) \sin \phi}{(1 + r_{j'}) \cos \phi}\right]\\
\end{aligned}
\end{equation}
where we define the leakage ratio $r \equiv q/p$. We see that for general $\phi$ the coherence and phase lag estimators are no longer consistent, a bias induced by spectral leakage is introduced. The magnitude of this bias is controlled by $\phi$ and the leakage ratio, which is a function of the shape of the input spectra. In Figure \ref{fig:leakage}, we show the effect of this leakage compared to sample input simulation parameters and show our derivation accurately predicts these effects. 

\begin{figure}
    \centering
    \includegraphics[width=0.95\linewidth]{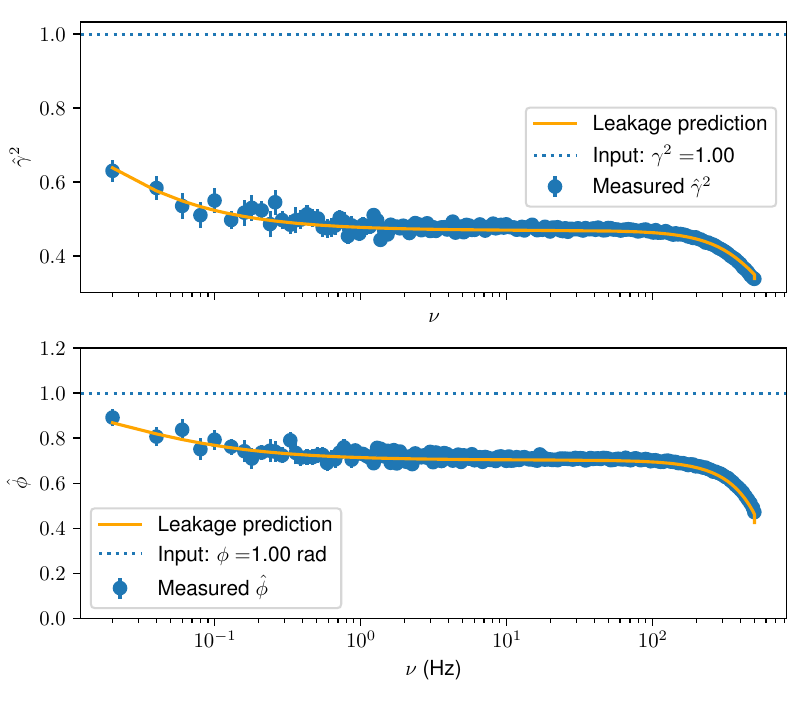}
    \caption{Effects of spectral leakage on a simulation with input parameters $\coh = 1.0$, $\phi = 1.0$ rad, $P_X \propto \nu^{-2}$. Recovered coherence and phase lag values are shown after segmenting $10,\!000$ second synthetic light curves into 200 segments and calculating Fourier properties from the average of the segments. The predicted effects of leakage (orange lines, equations (\ref{eq:leaky_coh}) and (\ref{eq:leaky_phi})) match the observed values (blue points).}
    \label{fig:leakage}
\end{figure}

\subsubsection{White Noise Spectra}
For the specific case of a white noise spectrum, $P_X = 1$, we can derive an approximate closed form for $r$. In this specific case we can write 
\begin{equation}
    q_{j'}^{\text{white}} = \frac{1}{N^2} \sum_{k = 1}^{N/2 - 1} | W_{N-k,j'}|^2.
\end{equation}
We will approximate $q_{j'}^{\text{white}}$ in the large $N$ limit by considering an integral instead of the discrete sum. Via a change of variables, $u = \frac{k}{N} + \frac{j'}{N_s}$, we write
\begin{equation}
    \begin{aligned}
        q_{j'}^{\text{white}} &\approx \frac{1}{N} \int_{1/N + j'/N_s}^{1/2 +j'/N_s} \frac{\sin^2(N_s \pi u)}{\sin^2 (\pi u)} du.\\ 
    \end{aligned}
\end{equation}
We approximate this integral replacing its rapidly oscillating part $\sin^2(N_s \pi u)$ with its average over the domain of integration. We write
\begin{equation}
    \begin{aligned}
        q_{j'}^{\text{white}} &\approx \frac{1}{2N} \left( 1 + \frac{N \sin(\frac{2 N_s \pi}{N})}{N_s \pi (N-2)}\right) \int_{1/N + j/N_s}^{1/2 +j/N_s} \frac{1}{\sin^2 (\pi u)} du.\\ 
        &= \frac{1}{2\pi N} \left( 1 + \frac{N \sin(\frac{2 N_s \pi}{N})}{N_s \pi (N-2)}\right) \\
        &\qquad \left( \cot \left (\frac{\pi}{N} + \frac{j'\pi}{N_s}\right) + \tan \left( \frac{j'\pi}{N_s}\right)\right)\\
        &\approx \frac{1}{\pi N} \csc \left( \frac{2 \pi j'}{N_s}\right),
    \end{aligned}
\end{equation}
where we take the large $N$ limit in the final step, keeping only terms to $O(1/N)$. In lieu of calculating $p$ directly, we exploit Parseval's theorem to note that the total kernel power must be the same in the time and frequency domains
\begin{equation}
    \begin{aligned}
        p^{\text{white}}_{j'} + q^{\text{white}}_{j'} &= \frac{1}{N^2}\sum_{k =0}^{N-1} |W_{k,j'}|^2 \\
        &= \frac{1}{N} \sum_{n=0}^{N_s-1} \left| e^{-2\pi i n j'/N_s} \right|^2 \\
        &= \frac{1}{N} \sum_{n=0}^{N_s-1} 1 \;=\; \frac{N_s}{N} \\
        p^{\text{white}}_{j'} &\approx \frac{N_s}{N} - \frac{1}{\pi N} \csc \left( \frac{2 \pi j'}{N_s}\right).
    \end{aligned}
\end{equation}
Finally then, 
\begin{equation}
    r_{j'}^{\text{white}} \approx \frac{\csc \left( \frac{2 \pi j'}{N_s} \right)}{\pi N_s - \csc \left( \frac{2 \pi j'}{N_s}\right)}
\end{equation}

\subsection{Dependence on Spectral Redness}
The closed form for $r^{\text{white}}_{j'}$ derived above corresponds to the limit in which $P_X$ is uniform across the band. For non-uniform spectra, $r_{j'}$ depends on how power is distributed across the frequencies of the long time series. Redder spectra, those that concentrate power at lower frequencies, experience larger leakage effects. The structural reason is visible directly in the definitions of $p_{j'}$ and $q_{j'}$. The kernel $|W_{k,j'}|^2$ that defines $p_{j'}$ has a peak at the matching long-series frequency $k_0 = j' N/N_s$, where it attains the value $N_s^2$. The conjugate kernel $|W_{N-k,j'}|^2$ that defines $q_{j'}$ has no such peak for $k \in [1, N/2-1]$ and $j' \in [1, N_s/2-1]$, and the sum is fed entirely by sidelobe leakage from across the band, weighted by $P_{X,k}$. For a sufficiently red $P_X$, the integrated power in the lowest few bins of the long-series spectrum is large enough that the sidelobe contribution to $q_{j'}$ becomes comparable to the local main-lobe contribution to $p_{j'}$, particularly at high $j'$, where the main lobe samples a small $P_{X,k_0}$.

The application in Section~\ref{sec:results} uses Lorentzian power spectra rather than pure power laws. A Lorentzian decays as $\nu^{-2}$ outside its peak but carries no diverging power toward DC, so its low-frequency contribution to $q_{j'}$ remains bounded. The recovered coherence and lag profiles in Fig.~\ref{fig:usage_example} are minimally affected by leakage.

\subsection{Mitigation}
If one wishes to simulate synthetic time series which do not display the effects of spectral leakage, mitigation options are available. While it is necessary to average the cross spectrum to measure the coherence function, one can achieve this by averaging neighboring frequency values instead of segmenting generated timeseries and averaging segments. Working with unsegmented light curves is sometimes done in AGN studies \citep[see e.g.,][]{kara2016}. While not commonly done in X-ray timing work, various alternate kernels which minimize the magnitude of sidelobes could reduce the level of leakage. Such kernels are more commonly used in broader signal processing work \citep[for a catalog of common kernels see][]{nuttall1981}.

Previous works have discussed the statistical properties of power spectra under the effects of spectral leakage \citep{deeter1982, deeter1984} and recommended methods of mitigation \citep{papadakis1993,zhu2016, epitropakis2016, epitropakis2017}. Applications of these methods to synthetic data could produce synthetic timeseries which better match their input properties. Complementarily, the method described in this paper can be used to estimate the level of leakage observed and test various mitigation methods by providing a known ground truth. 
\end{edit-sec}


\bsp	
\label{lastpage}
\end{document}